# SETI is Part of Astrobiology

Jason T. Wright

Department of Astronomy & Physics
Center for Exoplanets and Habitable Worlds
Penn State University

Phone: (814) 863-8470

astrowright@gmail.com

# I. SETI is Part of Astrobiology

"Traditional SETI is not part of astrobiology" declares the NASA Astrobiology Strategy 2015 document (p. 150). This is incorrect.[1]

Astrobiology is the study of life in the universe, in particular its "origin, evolution, distribution, *and future* in the universe." [emphasis mine] Searches for *biosignatures* are searches for the results of interactions between life and its environment, and could be sensitive to even primitive life on other worlds. As such, these searches focus on the origin and evolution of life, using *past* life on Earth as a guide.

But some of the most obvious ways in which Earth is inhabited *today* are its *technosignatures* such as radio transmissions, alterations of its atmosphere by industrial pollutants, and probes throughout the Solar System. It seems clear that the *future* of life on Earth includes the development of ever more obvious technosignatures. Indeed, the NASA Astrobiology Strategy 2015 document acknowledges "the possibility" that such technosignatures exist, but erroneously declares them to be "not part of contemporary SETI," and mentions them only to declare that we should "be aware of the possibility" and to "be sure to include [technosignatures] as a possible kind of interpretation we should consider as we begin to get data on the exoplanets."

In other words, while speculation on the nature of biosignatures and the design of multi-billion dollar missions to find those signatures is consistent with NASA's vision for astrobiology, speculation on the nature of technosignatures and the design of observations to find them is not. The language of the strategy document implies NASA will, at best, tolerate its astrobiologists considering the possibility that anomalies discovered in the hunt for biosignatures might be of technological origin.

But there is no *a priori* reason to believe that biosignatures should be easier to detect than technosignatures—indeed, we have had the technology to detect strong extraterrestrial radio signals since the first radio SETI searchers were conducted in 1959, and today the scope of possibly detectable technosignatures is much larger than this. Furthermore, intelligent *spacefaring* life might spread throughout the Galaxy, and so be far more ubiquitous than new sites of abiogenesis. Life might be much easier to find than the NASA strategy assumes.

Indeed it has been cynically, but not untruthfully, noted that NASA eagerly spends *billions* of dollars to search for "stupid" life passively waiting to be found, but will spend almost nothing to look for the intelligent life that might, after all, be trying to get our attention. This is especially strange since the discovery of *intelligent* life would be a much more profound and important scientific discovery than even, say, signs of photosynthesis on Ross 128*b.*

---

[1] Indeed, broad swaths of the astrobiology community disagree with NASA's assertion. For instance, SETI was included as a component of astrobiology in The Astrobiology Primer v.2.0 (Domagal-Goldman & Wright 2016), and SETI activities fall under the Carl Sagan Center for astrobiology at the SETI Institute (which, despite the name, conducts a broad range of science, including many sub-fields of astrobiology).

Further, since technosignatures might be both *obvious* and *obviously artificial* SETI also provides a shortcut to establishing that a purported sign of life is not a false positive, a major and pernicious problem in the hunt for biosignatures. SETI thus provides an alternative and possibly more viable path to the discovery of alien life than is reflected in NASA's astrobiology roadmap. Indeed, this was recognized explicitly in the panel reports of the Astro2010 decadal survey:

> *Of course, the most certain sign of extraterrestrial life would be a signal indicative of intelligence. [A radio] facility that devoted some time to the search for extraterrestrial intelligence would provide a valuable complement to the efforts suggested by the PSF report on this question. Detecting such a signal is certainly a long shot, but it may prove to be the only definitive evidence for extraterrestrial life.* (p.454, Panel Reports—New Worlds, New Horizons in Astronomy & Astrophysics)

## II. Why is SETI Neglected in NASA's Astrobiology Portfolio?

While it is not completely clear *why* NASA does not include SETI in its astrobiology portfolio, there are several factors that seem likely to be at play.

The first is the risk of public censure: SETI sometimes suffers from a "giggle factor" that leads some to conflate it with "ufology" or campy science fiction. Indeed, such an attitude likely led to the cancelation of the last NASA SETI efforts in the early 1990's, after grandstanding by US senators denouncing "Martian hunting season at the taxpayer's expense" (Garber 1999). Such attitudes harm all of science, and the National Academies should be clear that such a "giggle factor" must not be allowed to influence US science priorities.

The second is the erroneous perception that SETI is an all-or-nothing proposition that yields no scientific progress unless and until it succeeds in detecting unambiguous signs of interstellar communication. On the contrary, even with scant funding, SETI has historically been involved in some of the most important discoveries in astrophysics. Not only have the demands of radio SETI led to breakthroughs in radio instrumentation (see, for instance, the new Breakthrough Listen backend at the Green Bank 100-meter telescope, with bandwidth of up to 10 GHz, an ideal Fast Radio Burst detection device; Gajjar et al. 2017), but some of the most famous SETI false positives have proven to be new classes of astrophysical phenomena, including active galactic nuclei (CTA-21 and CTA-102, Kardashev 1964), pulsars (originally, if somewhat facetiously, dubbed "LGM" for "Little Green Men"), and perhaps the still-not-fully-understood "Tabby's Star" (KIC 8462852, Boyajian et al. 2016, Wright et al. 2016, Wright & Sigurdsson 2017).

Indeed, exactly because SETI seeks signals of obviously artificial origin, it must deal with and examine the rare and poorly understood astrophysical phenomena that dominate its false positives. Anomalies discovered during searches for pulsed and continuous laser emission (Howard et al. 2007, Wright et al. 2014, Tellis & Marcy 2015, 2017) broadband radio signals, large artificial structures (Dyson 1960, Griffith et al. 2015, Wright et al. 2016), and other astrophysical exotica push astrophysics in new and unexpected directions. If there is a perception that SETI little more than the narrow search for strong radio carrier waves producing a long string of null results it is because historically there has been essentially no funding available for anything else.

Third, there is the erroneous perception that, since radio SETI as been active for decades, its failure to date means there is nothing to find. On the contrary, the lack of SETI funding means that only a tiny fraction of the search space open to radio SETI has been explored (Tarter et al. 2010). Indeed, Robert Gray has estimated that the total integration time on the location of the Wow! Signal (the most famous and credible SETI candidate signal to date) is less than 24 hours (see, for instance, Gray et al. 2002). That is, if there is a powerful, unambiguous beacon in that direction with a duty cycle of around one pulse per day, we would not have detected a second pulse yet. Other parts of the sky have even less coverage. The truth is, we only begun to seriously survey the sky even for radio beacons, and other search methods have even less completeness.

Fourth, there is the erroneous perception that SETI will proceed on its own without NASA support. Indeed, the 2015 NASA Astrobiology Roadmap claims that "traditional SETI is…currently well-funded by private sources."  Even setting aside the non sequitur of considering the amount of private philanthropic funding when assessing the merits of the components of astrobiology, this is not a fair description of the state of the field. While it is true that the Breakthrough Listen Initiative has pledged to spend up to $100 million over 10 years, in truth its spending has been far below that level, and it is focused on a small number of mature search technologies. Beyond this initiative, private benefactors have supported the SETI Institute's Allen Telescope Array, but not at the level necessary to complete the array or fund its operations.

Fifth, there is the erroneous perception that the search for technosignatures is somehow a more speculative or risky endeavor than the search for biosignatures. We note that the entire field of astrobiology once faced a similar stigma. Chyba & Hand rebutted that perception in 2005:

> *Astro-physicists...spent decades studying and searching for black holes before accumulating today's compelling evidence that they exist. The same can be said for the search for room-temperature superconductors, proton decay, violations of special relativity, or for that matter the Higgs boson. Indeed, much of the most important and exciting research in astronomy and physics is concerned exactly with the study of objects or phenomena whose existence has not been demonstrated—and that may, in fact, turn out not to exist. In this sense astrobiology merely confronts what is a familiar, even commonplace situation in many of its sister sciences.*

Their rebuttal holds just as well as SETI today. Indeed, Wright & Oman-Reagan (2017) have articulated a detailed analogy between SETI and the relatively uncontroversial search for dark matter particles via direct detection. They argue that unlike with dark matter searches, with SETI, at least, we have the advantage that we *know* that the targets of our search (spacefaring technological species) arise naturally (because we are one).

Finally, there is an erroneous perception that SETI is exclusively a ground-based radio telescope project with little for NASA to offer. On the contrary, SETI is an interdisciplinary field (Cabrol 2016) and even beyond the potential for NASA's Deep Space Network to play an important role in the radio component of SETI, archival data from NASA assets have played an important role in SETI for decades: from Solar System SETI using interplanetary cameras, to waste heat searches using *IRAS* (Carrigan 2009) *WISE, Spitzer,* and *GALEX* (Griffith et al. 2015), to searches for artifacts with *Kepler* (Wright et al. 2016) and *Swift* (Meng et al. 2017). Future ground-based projects like LSST and space-borne projects like *JWST* and *WFIRST* will undoubtably provide additional opportunities SETI research both as ancillary output of legacy and archival programs and through independent SETI projects in their own right.

### III. Reinvigorating SETI as a Subfield of Astrobiology

One difficulty SETI faces is a negative feedback between funding and advocacy.

As it stands, SETI is essentially shut out of NASA funding. SETI is not mentioned at all in most NASA proposal solicitations, making any SETI proposal submitted to such a call unlikely to satisfy the merit review criteria. Worse, **the only mentions of SETI in the entire 2015, 2016, and 2017 ROSES announcements are under "exclusions,"** in the Exobiology section ("Proposals aimed at identification and characterization of signals and/or properties of extrasolar planets that may harbor intelligent life are not solicited at this time") and the Exoplanets section (as "not within the scope of this program.") In other words, SETI is ignored entirely in NASA proposal solicitations, *except for those most relevant to it*, in which cases it is *explicitly excluded*.

Meanwhile, other parts of astrobiology have flourished under NASA's aegis, which has incubated strategies for the detection of life elsewhere in the universe, and produced scientists who can advocate for mature roadmaps to the detection of life in the universe as part of NASA's astrobiology program. But now, twenty years after the last major NASA SETI program was cancelled, there are only a handful of SETI practitioners and virtually no pipeline to train more.

Thus there are only a few well-developed strategies to advocate for, and only a few scientists to advocate for them. This will doubtless be reflected in the number of white papers advocating SETI (like this one) versus those advocating other kinds of astrobiology responsive to the current call. This disparity should not be seen as indicating a lack of intrinsic merit of the endeavor of SETI, but as a sign of neglect of SETI by national funding agencies.

**Since SETI is, quite obviously, part of astrobiology, SETI practitioners should at the very least be *expressly encouraged* to compete on a level playing field with practitioners other subfields for NASA astrobiology resources.**

Doing so will uncork pent-up SETI efforts that will result in significant progress over the next 10 years and beyond. As a fully recognized and funded component of astrobiology, SETI practitioners will be able to develop new search strategies, discover new astrophysical phenomena and, critically, train a new generation of SETI researchers to guide NASA's astrobiology portfolio to vigorously pursue the discovery of all kinds of life in the universe—both "stupid" *and* intelligent.

And if, as many suspect, technosignatures prove to be closer to our grasp than biosignatures, then including of SETI in NASA's astrobiology portfolio will ultimately lead to one of the most profound discoveries in human history, and a reinvigoration of and relevance for NASA not seen since the Apollo era. In retrospect, we will wonder why we were so reluctant to succeed.